\documentclass[12pt]{article}
\usepackage{epsf,cite,a4wide,latexsym}
\newcommand{\beq}{\begin{equation}}
\newcommand{\eeq}{\end{equation}}
\newcommand{\bea}{\begin{eqnarray}}
\newcommand{\eea}{\end{eqnarray}}
\newcommand{\bda}{\begin{eqnarray*}}
\newcommand{\eda}{\end{eqnarray*}}

\begin{document}

\vskip -4cm

\begin{flushright}
MIT-CTP-3244
\end{flushright}

\vskip
0.2cm

{\Large
\centerline{{\huge Monopole and vortex content of a meron pair}} }
\vskip 0.8cm

\centerline{ {\Large \'Alvaro Montero$^{(1)}$ and John W. Negele$^{(2)}$ } }

\vskip  0.4cm
\centerline{(1) Department de F\'{\i}sica Fonamental}
\vskip -0.15cm
\centerline{Universitat de Barcelona}
\vskip -0.15cm
\centerline{Barcelona 08028}
\vskip -0.15cm
\centerline{Spain}
\vskip -0.15cm
\centerline{e-mail: montero@ffn.ub.es}
\vskip 0.8cm

\centerline{(2) Center for Theoretical Physics}
\vskip -0.15cm
\centerline{Laboratory for Nuclear Science and Department of Physics}
\vskip -0.15cm
\centerline{Massachusetts Institute of Technology}
\vskip -0.15cm
\centerline{Cambridge, Massachusetts 02139}
\vskip -0.15cm
\centerline{USA}
\vskip -0.15cm
\centerline{e-mail: negele@mitlns.mit.edu}
\vskip 0.8cm

\begin{center}
{\bf
ABSTRACT}
\end{center}
We investigate the monopole and vortex content of a meron pair by calculating
the points at which the transformation to the Laplacian Center Gauge is ill-defined 
and by studying the behavior of Wilson loops.
These techniques reveal complementary aspects of the vortex and monopole
structure, including the presence  of closed monopole lines and closed vortex
surfaces joining the two merons, and evidence for intersecting vortex
surfaces at each meron.


\newpage

\section{Introduction}

The QCD vacuum is characterized by two striking phenomena, the breaking of
chiral symmetry and the confinement of color charge. Chiral symmetry breaking
may be understood in terms of localized topological excitations of the
gluon field and their associated quark zero-modes that produce a non-vanishing
value of the chiral condensate.
Classical instanton \cite{instanton} solutions of the Yang Mills equations with 
topological charge $Q=1$ and their quantum fluctuations provide a
physical foundation for these topological excitations and thus a natural
understanding of chiral symmetry breaking.

In contrast, the mechanism for confinement is not presently well
understood, and various pictures have been investigated to try to
explain it in  terms of relevant structures in the QCD vacuum.
Various point-like solutions to the Yang Mills equations, which fall off
at large distances in all space-time dimensions, have been considered.
Although Q=1 instanton solutions provide an understanding of chiral
symmetry breaking, in the dilute gas and instanton liquid approximations
they do not lead to confinement\cite{chen}. Merons, topological charge
$\frac{1}{2}$ solutions found by De Alfaro, Fubini and Furlan \cite{meron},
are more strongly disordering  objects than instantons and were proposed as
a mechanism for confinement by Callan, Dashen and Gross \cite{Callan:1977qs}. 
Fractons, also solutions of the Yang Mills equations of motion with fractional 
topological charge, appear on the four-dimensional torus, $T^4$, when twisted 
boundary conditions are imposed \cite{thoof}. The possible relevance of these 
objects to confinement was pointed out in \cite{thoof2}, and a scenario for
confinement based on the fractional charge solution found in reference
\cite{q1d2}, was proposed by Gonz\'alez-Arroyo and Mart\'{\i}nez \cite{invest}.

One and two-dimensional structures in the QCD vacuum have also been considered
as mechanisms for confinement. In the dual superconductor picture \cite{monopoles}, 
the condensation of monopoles in the QCD vacuum leads to confinement.
Monopoles are one-dimensional curves in space-time that appear in QCD as
defects in the abelian gauges proposed by 't Hooft \cite{abproj}. The gauge is
fixed up to the Cartan subgroup of the gauge group and monopoles appear at
points in space where this gauge fixing is ill-defined, leaving a gauge
freedom larger than the abelian subgroup. In the  vortex theory\cite{vortices},
confinement is due to the condensation of vortices. Vortices are two-dimensional
surfaces carrying flux in the center of the SU(N) group, which means that
a Wilson loop intersecting the surface of the vortex takes the value of one of
the elements of the center of the group. Classical vortex solutions to
the SU(N) Yang Mills equations have been found numerically \cite{montero}.

The mechanism for chiral symmetry breaking and the alternative descriptions of
confinement are not mutually exclusive - rather they are highly interrelated. The
fact that the intersection of two vortices has topological charge $\frac{1}{2}$
\cite{Engelhardt,jahn,Cornwall} provides a provocative connection between chiral
symmetry breaking and confinement and suggests that the confinement
properties of charge $\frac{1}{2}$ merons may also be understood in terms
of the intersections of  vortices. In addition, as elaborated below, monopole
lines lie on vortex surfaces, so that both structures coexist and may be studied
simultaneously. In this picture, a meron pair corresponds to the intersection of 
two closed vortex sheets containing closed monopole loops and provides the
simplest system in which one could explore this structure quantitatively.
As the separation between the merons decreases to zero and they merge into an
instanton, one would expect a vortex sheet and a monopole loop on it to
shrink to a point at the center of the instanton \cite{brower,Bruckmann}. A 
similar picture of the separation of an instanton into two fractionally charged 
objects connected by hedgehog world lines is given in reference \cite{Cornwall2}.

In this article we investigate numerically the monopole and vortex content of
a meron pair in SU(2) Yang Mills theory by calculating the points at
which Laplacian Center Gauge fixing is ill-defined\cite{forcrand,deflcg}
and by calculating the behavior of Wilson loops. The monopole and vortex
content of an isolated meron has already been studied analytically by
Reinhardt and Tok \cite{Reinhardt} using Laplacian Center Gauge fixing and
Wilson loops, and provides an essential foundation for the present work.
Since their work, as well as that of others, has shown Laplacian Center
Gauge fixing to be an imperfect tool, in this study we also explore the
limitations of this tool as well as the physics  of the QCD vacuum.

The outline of this letter is the following. In section 2 we describe the
meron pairs that we study and in section 3 we use Wilson loops to explore
their vortex content. Section 4 presents the monopole and vortex
content of these configurations determined from Laplacian Center Gauge defects
and section 5 summarizes our conclusions.

\section{The meron pair}

Merons \cite{meron} are solutions to the classical Yang-Mills equations of motion
in four Euclidean dimensions, which can be written as
\begin{equation}
A_\mu^a(x) =  \eta_{a\mu\nu} \frac{x^\nu}{x^2} \ ,
\label{general}
\end{equation}
where $\eta_{a\mu\nu}$ is the 't Hooft symbol. Using the conformal symmetry
of the classical Yang-Mills action, it can be shown that in addition to a
meron at the origin, there is a second meron at infinity, and these two merons
may be mapped to arbitrary positions. The gauge field for the two merons
\cite{meron} is
\begin{equation}
A_\mu^a(x) =  \eta_{a\mu\nu}
\left[ \frac{x^\nu}{x^2}+
\frac{(x-d)^\nu}{(x-d)^2} \right] \ .
\label{conform}
\end{equation}

This gauge field for the meron pair has infinite action density at points
$x_\mu=\{0,\, d_\mu\}$. To avoid the problem of these singularities, we use
the following expression
\cite{Callan:1977qs}
\begin{equation}
A_\mu^a(x) =  \eta_{a\mu\nu} x^\nu
\left\{
\begin{array}{cll}
\displaystyle
\frac{2}{x^2+r^2} \ ,\qquad &
\sqrt{x^2} < r \ ,
\\[3ex]
\displaystyle
\frac{1}{x^2} \ ,\qquad & r <
\sqrt{x^2} < R \ ,
\\[3ex]
\displaystyle
\frac{2}{x^2+R^2} \ ,\qquad & R <
\sqrt{x^2} \ .
\end{array}
\right.
\label{caps}
\end{equation}
Here, the singular meron fields for $\sqrt{x^2}<r$ and $\sqrt{x^2}>R$
are replaced by instanton caps, each containing topological charge $\frac{1}{2}$
to agree with the topological charge carried by each meron.
We study the monopole and vortex content of this configuration by putting
the gauge field on a lattice of size $N_t \! \times \! N_s^3$. For details of
the procedure for putting the meron pair on the lattice and relaxing it to a
solution of the field equations, see reference \cite{Negele:2000}.

In this article, we analyze four meron pair configurations obtained on
$N_t\!\times\!N_s^3$ lattices with $N_s\!=\!16,24$ and $N_t\!=\!2N_s$.
We study configurations with different cap sizes, $c$, distances between
merons, $d$, and sizes of the lattice, $N_s$. We used a configuration with
$N_s\!=\!16$, $c\!=\!4$ and $d\!=\!10$ (configuration I), and three configurations
with $N_s\!=\!24$: one with $c\!=\!1$ and $d\!=\!12$ (configuration II),
one with $c\!=\!5$ and $d\!=\!12$ (configuration III), and one with
$c\!=\!1$ and $d\!=\!16$  (configuration IV). We have checked that the field
strength from each of the lattice configurations has  the  essential properties
described in reference \cite{Negele:2000} for the continuum field strength. We
have also applied up to five cooling sweeps to the meron pair configurations in 
order to relax them close to lattice solutions, and checked that the monopole 
and vortex content for these meron pair configurations are independent of this 
cooling. Although we do not explicitly address Dirac zero modes in this work, 
note that the zero mode for a meron pair configuration has been calculated for 
a range of separations in reference \cite{Negele:2000} and displays two peaks 
at the positions of the merons.

\section{Vortex content from Wilson loops}

Before considering Laplacian Center Gauge fixing, it is useful to describe  
the vortex content obtained from calculating Wilson loops. For a single, singular 
meron at the origin, it has been  shown that a circular Wilson loop around the 
origin in any of the six planes defined by a pair of coordinate axes ($x,y,z,t$)
has the value $-1$ for any size of  the circle \cite{Reinhardt}. Hence, Wilson 
loops indicate the presence of a vortex  surface on all the planes defined by pairs 
of coordinate axes.
For our configurations, regularized meron pairs, we studied two sets of Wilson
loops. For the $xy$, $xz$ or $yz$ plane, planes orthogonal to the line joining both
merons, we calculated a square Wilson loop of size $r \! \times r$ with one of the
merons in  the center of the loop. The results for configurations II and III
(distance between  merons $d=12$ and cap sizes $c=1,5$), and for the $xy$ plane, 
are shown in figure 1A. We see that at short distance, the value of the Wilson 
loop goes from  $+1$ towards the value $-1$,  as for a single meron, and only 
changes this behavior at large distance where the contribution of the second meron
starts to be significant, approaching the value $+1$ when the loop is bigger than 
the distance between the merons. We also see in figure 1A the effect of the cap size. 
The cap gives a characteristic size $c$ to the meron, which is reflected in the 
distance one must go for the value of the Wilson loop to start to be approach $-1$ 
and thus enclose the vortex flux. Note that for the original singular meron, this 
size would be zero. The results obtained for the other two planes, $xz$ and $yz$, 
are the same, showing the underlying spherical symmetry in the spatial directions 
of the meron  pair. Results for the other two configurations were completely
analogous, with the curves simply reflecting the corresponding cap sizes
and separation between merons.

For the $xt$, $yt$ or $zt$ planes, which include the line joining both
merons,  we calculated a rectangular Wilson loop of size $r \! \times \! 2r$ 
with one of the merons in the center of the loop. The results in the
$xt$ plane for the same configurations as in figure 1A, are shown in figure 1B. 
We see that again at short distances, the Wilson loop goes from $+1$ to
$-1$ as the size of the loop increases, and as $r$ exceeds half the separation
between merons, the Wilson loop begins to approach +1, which it will reach when
both merons are included. Again, the loop must be larger than the cap size, $c$, 
to enclose all the vortex flux.  As before, the results for the other two planes, 
$yt$ and $zt$, are the same, and  the other configurations show analogous behavior
reflecting the other cap sizes and separations.

\begin{figure}
\caption{ {\footnotesize Wilson loops for a meron pair. Figure A shows the
values of $r \times r$ Wilson loops in the $xy$ plane centered at the
maximum of one of the  merons as a function of $r$ for configurations
with separation $d\!=\!12$ and cap sizes $c$ = 1 and 5 (configurations II and III). 
Figure B shows the values of $r\times 2r$ Wilson  loops in the $xt$ plane centered 
at the maximum of one  of the  merons as a function of $r$ for the same configurations
as in figure A. In both figures, lines are plotted joining the calculated points to 
guide the eye.
          }
        }
\vbox{ \hbox{  \vbox{ \epsfxsize=3.0truein \epsfysize=3.25truein
                      \hbox{\epsffile{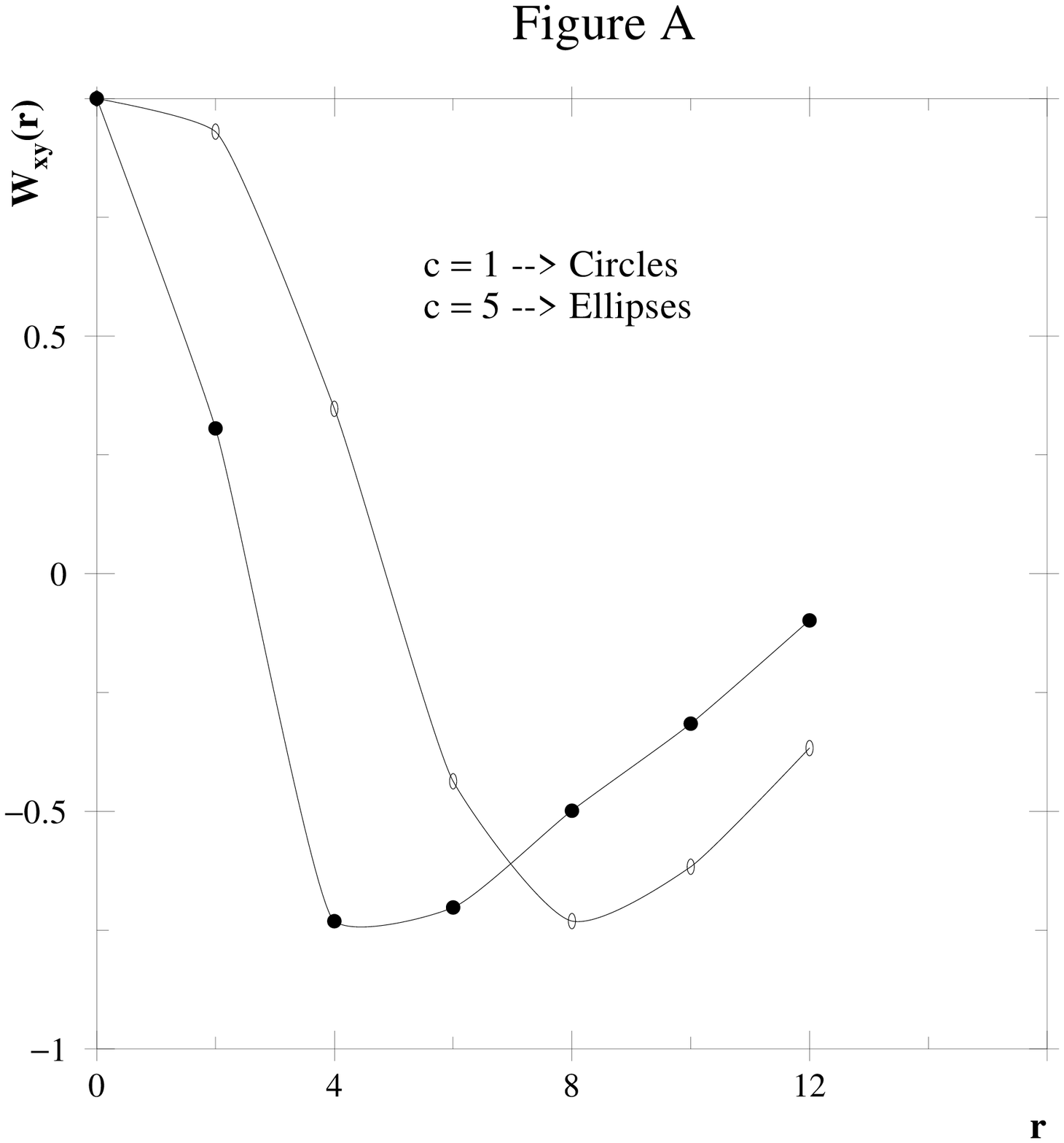} } }
        \hfill \vbox{ \epsfxsize=3.0truein \epsfysize=3.25truein
                      \hbox{\epsffile{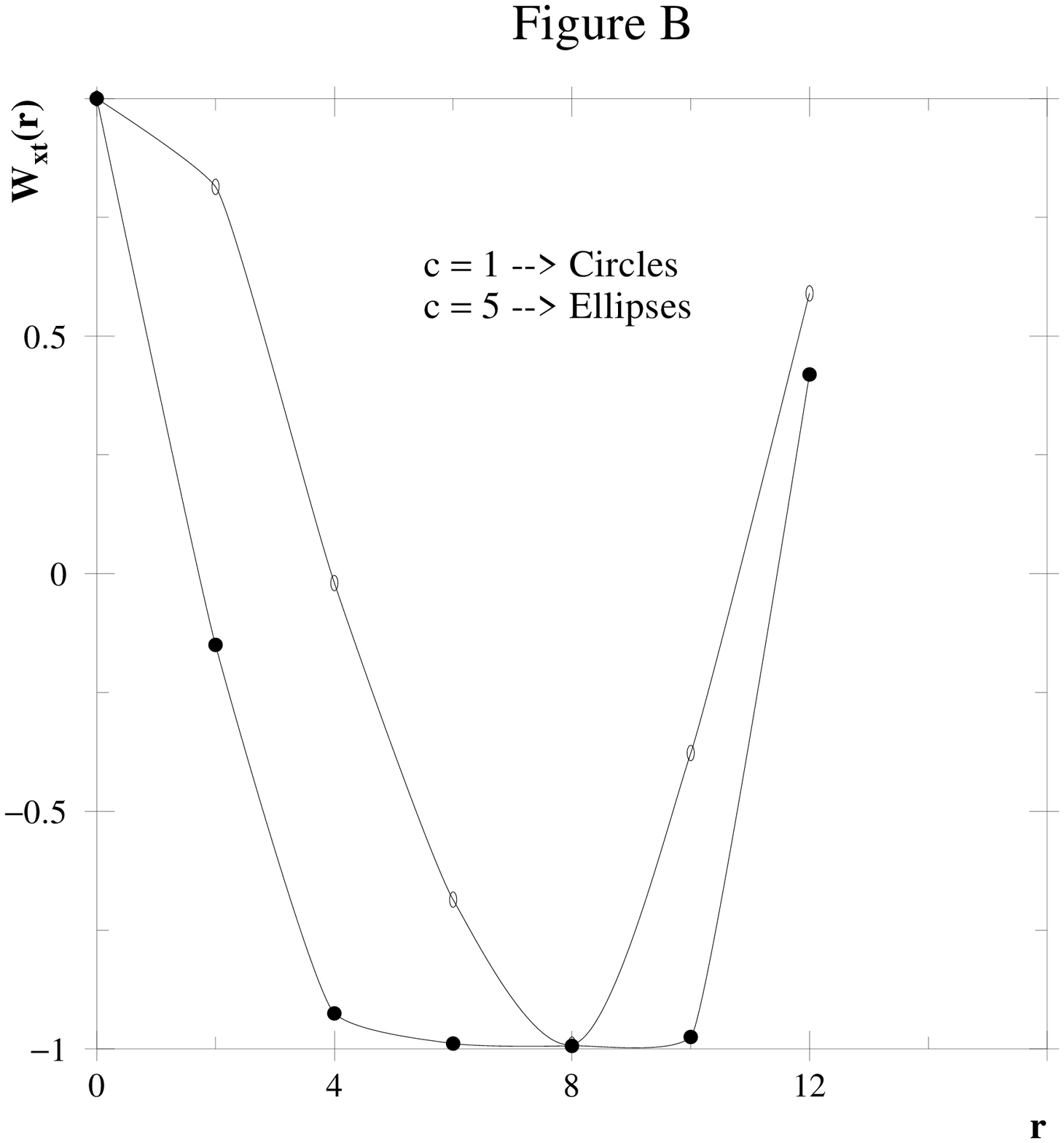} } }
            }
}
\end{figure}

The conclusion from this study of Wilson loops in a meron pair is that, like an
isolated meron, a meron in a pair behaves like a source or sink for flux in
non-trivial elements of the center of the group for all six planes defined
by the Cartesian axes, and the size of the source or sink is of order of the cap
size, $c$. Thus, each meron corresponds to the intersection of orthogonal pairs of
vortices.

\section{Monopole and vortex content from \\ Laplacian
Center Gauge defects}

In this section, we present the monopole and vortex content of the meron pair
configurations described in the previous section, as inferred from the
points at which gauge fixing to the Laplacian Center Gauge is ill-defined.

Fixing the gauge to Laplacian Center Gauge \cite{forcrand,deflcg} involves 
the use of the two eigenvectors with lowest eigenvalues, $\psi^a_1(n)$ and 
$\psi^a_2(n)$, of the Laplacian operator,
\beq
{\cal L}_{nm}^{ab}(R) = \sum_{\mu} \left( 2 \delta_{nm}
\delta^{ab} -
R^{ab}(n,\mu)
\delta_{m,n+\hat{\mu}}  - R^{ba}(m,\mu)
\delta_{n,m+\hat{\mu}} \right)
\label{eq:oper}
\eeq
in the presence of a gauge field $R^{ab}(n,\mu)$ in the adjoint representation
of the gauge group. The lowest eigenvector, $\psi^a_1(n)$, is rotated to
the ($\sigma_3$) direction in color space. This step fixes the gauge up to
the abelian subgroup of the SU(2) group. The U(1) abelian freedom is fixed
by imposing the additional condition that the $\psi^a_2(n)$ eigenvector
is rotated to lie in the  positive ($\sigma_1,\sigma_3$) half-plane.
After these two steps, the gauge is completely fixed up to the center
degrees of freedom.

Monopoles and vortices are found in Laplacian Center Gauge as defects
of the gauge fixing procedure, which means we have to look at the points
at which the gauge fixing prescription is ill-defined. The
first step, rotation of the first eigenvector to the third direction in
color space, is ill-defined if $\psi^a_1(t,x,y,z)=0$.  This defines lines
in four-dimensional space and these lines are identified  as monopole lines. 
The second step, rotation of the second eigenvector to the positive
$(\sigma_1,\sigma_3)$ half-plane, is ill-defined at points at
which the first and second eigenvectors are parallel. This condition
defines surfaces in four dimensional space and these surfaces are
identified as vortex sheets.

To fix to the Laplacian Center Gauge we use the algorithm presented in 
\cite{conjgrad} to calculate the lowest eigenvectors of the Laplacian operator.  
We calculate the four eigenvectors with lowest eigenvalues, and find that the three
lowest eigenvalues are degenerate. With two vectors chosen from these three, or
from linear combinations of these three, we can fix the gauge to Laplacian
Center Gauge. Note that because of the degeneracy in the lowest eigenvalues, the
monopole and vortex content is ambiguously defined, and in this work we will
consider all the different monopole and vortex  patterns that may be obtained
from the lowest eigenvectors.

Before considering the monopole and vortex content of our meron pair
configurations, it is useful to review the monopole and vortex content of two
limiting cases,  an instanton and a single meron. The eigenfunctions of the
lowest state of the  Laplacian for these two cases are known 
analytically\cite{Reinhardt}. For an isolated instanton  there are
three degenerate eigenfunctions, a monopole is only be located at the
origin since it is the only point at which the eigenfunctions vanish, and
there are no vortices because the three eigenfunctions are always mutually
orthogonal. For the  single meron, there are four degenerate eigenfunctions
and the monopole content  depends on the choice of lowest eigenvector. For
the functions given in \cite{Reinhardt} it is easy to see that the $i^{th}$
eigenvector has a monopole line along the $i^{th}$ coordinate axis. We may
think of this monopole line as joining the meron at the origin with the second
meron at infinity, and the different lines arising from the different eigenvectors
or combinations of them simply reflects the fact that the second meron may be
reached at any position on the sphere at infinity. The vortex content also depends
on the choice of  the two lowest eigenvectors. It is easy to see that if one takes
the $i^{th}$ and $j^{th}$  eigenvectors, the vortex content is given by the plane
generated by the $i^{th}$ and $j^{th}$  coordinate axis. Taking other combinations 
of these four eigenvectors produces more complicated  results, like one of the 
examples presented in \cite{Reinhardt}, in which one obtains three vortex sheets
given by three planes intersecting at the origin. It is noteworthy that this
construction never generates the expected geometry of two intersecting vortices 
which in turn contain monopole lines, revealing that the Laplacian Center Gauge 
defects do not provide a completely satisfactory picture even in this analytically
solvable case.

We now consider the monopole and vortex content of our meron pair
configurations extracted from the three degenerate eigenvectors of the
Laplacian. We will show that there is a vortex surface that looks like an 
ellipsoid of revolution touching the center of each meron at each tip, and that the
monopole loops lie on this surface.  The zeros in figures 2B and 2C correspond to
longitudinal and axial sections of this surface respectively.
The first quantity we examine is the modulus $\Psi(n) \! = \!
\sqrt{\sum_{a=1}^3 (\psi^a(n))^2}$ and we look for points at which
$\Psi(n) \! = \! 0$.  One of the problems we have to face is the different
monopole pictures  obtained by choosing different eigenvectors or a linear
combinations thereof. We have looked for a combination in which the monopole
content is particularly clear and found that it was useful to  minimize the
sum of the modulus in a specified plane by taking linear combinations of the
three eigenvectors.  Doing this, we have found a combination in which the
first eigenvector, I,  has a monopole loop in the $xt$ plane joining the two
merons, the second  eigenvector, II, has the same monopole loop but in the 
$yt$ plane and the third eigenvector,  III, has the same  monopole loop in 
the $zt$ plane. A picture of the loop coming from eigenvector I, the loop in 
the $xt$ plane, can be seen in figure 2B, which is explained in more detail  
below. The vortex content obtained from eigenvectors I, II and III is also 
quite clear. If we choose eigenvectors I and II as the two lowest 
eigenvectors, the vortex surface looks like an ellipsoid of
revolution around the $t$ axis in the $xyt$ coordinates, with this surface
including the monopole loop in  the $xt$ plane coming from eigenvector 
I and the monopole loop in the $yt$ plane coming from  eigenvector II. If 
we choose eigenvectors I and III as the two lowest eigenvectors, the vortex 
surface is the same but in this case in the $xzt$ coordinates and if we 
choose  eigenvectors II and III we see the same surface in the $yzt$ 
coordinates.

\begin{figure}
\caption{ {\footnotesize Figure A shows the action density S(t,x,y,z) for
the meron pair with $d=16$ and $c=1$ (configuration IV) as a function of $x$ and $t$,
with $y$ and $z$ fixed to the values that maximize the action density. Figure B shows 
the absolute value of the discriminant of the three lowest Laplacian eigenvectors,
$D(t,x,y,z)$, to the $1/4$ power  as a function of $x$ and $t$,
for the  same meron pair configuration and  values of  the $y$ and $z$
coordinates used in figure A. Figure C shows the absolute value of
$D(t,x,y,z)$ to the $1/4$ power as a function of $x$ and $y$ for $z$ fixed
to the value that maximizes the action density and $t$ fixed to the midpoint
between the two merons. } }
\vbox{ \hbox{  \vbox{ \epsfxsize=2.1truein \epsfysize=2.1truein
                      \hbox{\epsffile{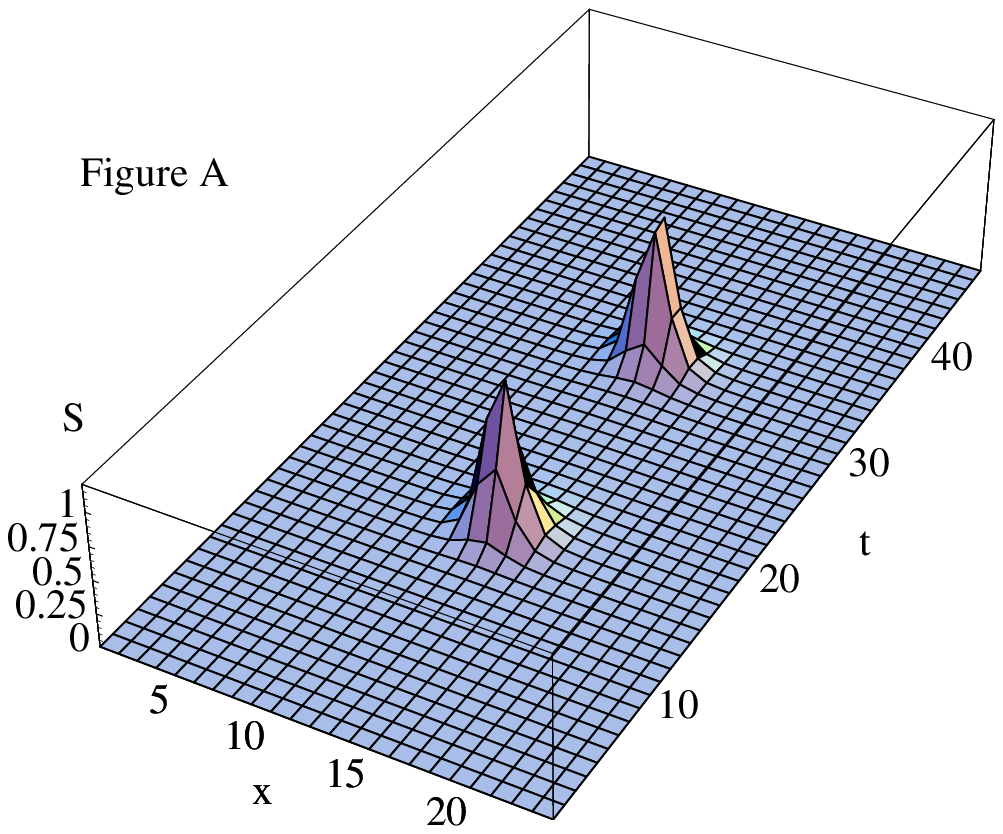} } }
        \hfill \vbox{ \epsfxsize=2.1truein \epsfysize=2.1truein
                      \hbox{\epsffile{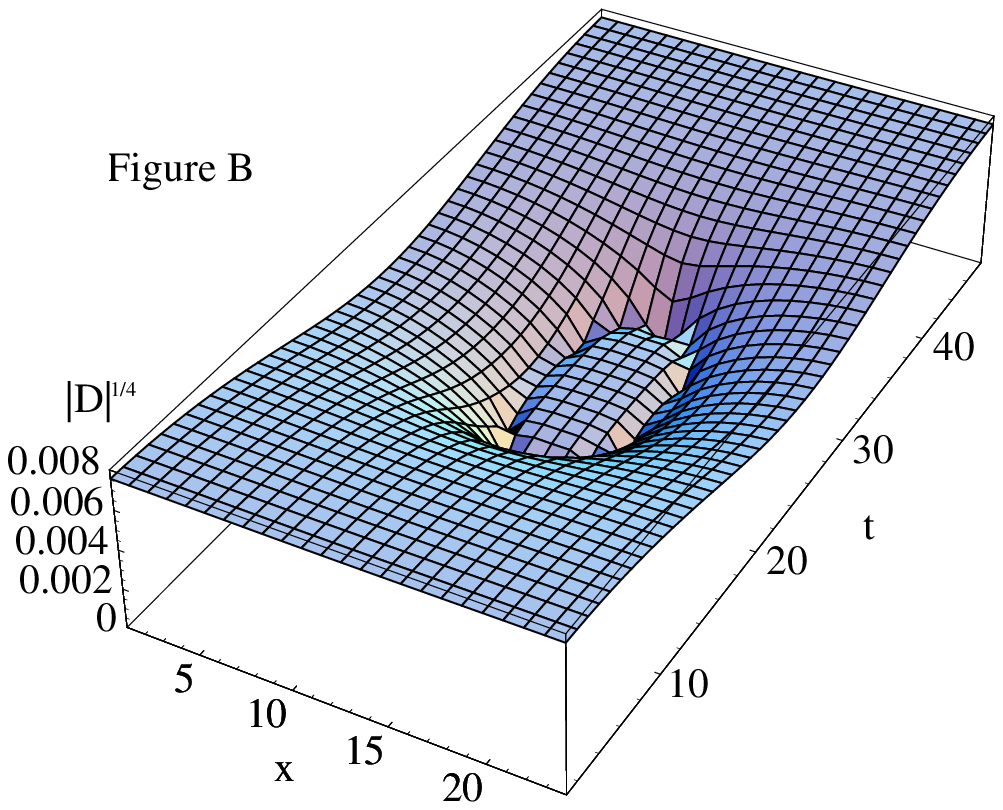} } }
        \hfill \vbox{ \vspace{0.5 cm} \epsfxsize=1.7truein
\epsfysize=1.7truein
                      \hbox{\epsffile{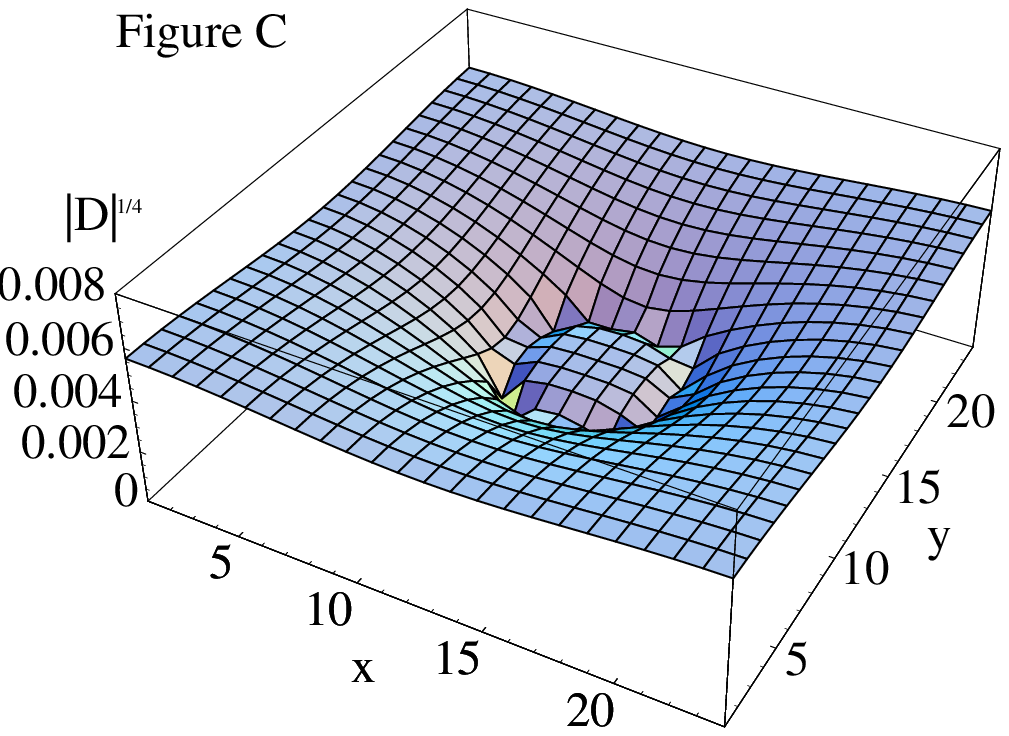} } }
            }
     }
\end{figure}

To obtain the locus of all the points which can be monopoles or vortices for our
meron configurations, we calculate  the determinant of these three vectors,
I, II and III, at each lattice point. First, note that if any of the vectors is
zero, the condition to find monopoles,  the determinant is zero. Second, note that
if there is a linear combination  between them giving a zero vector, the condition to
find vortices, the determinant is also zero. Finally, note that the determinant is
independent under linear combinations of the three vectors. Hence, all the points
for our meron configurations that can be monopoles or vortices are determined
by the condition that the determinant vanishes.

The result we obtain is the following. We find a region on the lattice in  which 
the determinant is always positive and another one in which it is always negative,  
both regions separated by a three-dimensional volume in which the determinant
vanishes and defines all positions that can be monopoles or vortices. We describe 
the shape of this volume by showing some of its two-dimensional sections. First, we 
show its temporal dependence. Consider the determinant as a function of $x$ and $t$, 
for values of $y$ and $z$ fixed to the values that maximize  the action density for 
these coordinates. Figure 2B shows the  absolute value of the determinant (raised to  
the $1/4$ power to see the curve more clearly).  The curve on which the determinant 
vanishes is similar to an ellipse, and joins the two maxima of the action density in
the $x$, $t$ coordinates. To see that this loop joins the maxima in the action density
to within a fraction of a lattice spacing, we show in figure 2A the action density as 
a function of $x$ and $t$, and for the same fixed values of $y$ and $z$. This loop is 
the same as the monopole loop described above defined by  the  points at which the 
modulus of eigenvector I vanishes. If we look at the determinant as a function of $y$ 
and $t$ (or $z$ and $t$), and for  values of $x$ and $z$ (or $x$ and $y$) fixed to
the maxima  in the action density for these coordinates, we again obtain the same 
curves shown in figures 2A and 2B with  $x$ interchanged with  $y$ (or $z$). This 
curve in the $y$ and $t$ coordinates is  also the monopole loop defined by the points 
at which the modulus of eigenvector II vanishes (and the curve in the $z$ and $t$
coordinates is the monopole loop defined by eigenvector III).

Second, we show the spatial dependence for fixed time positions. If we label the 
temporal positions for the maxima in the action density for each meron as $t_1$ and 
$t_2$, we find that for values of the lattice position $t$ belonging to $[0,t_1]$ and 
$[t_2,N_t]$, the determinant is always positive for all points in the three-dimensional
lattice defined at each temporal point. For values of $t$ between the two merons, 
$t_1\!<\!t\!<\!t_2$, we find a spherical surface in which the determinant changes 
sign, and inside the sphere the determinant is negative. The radius of this sphere 
at each temporal point may be seen in figure 2B,  where it corresponds to half the 
width of the monopole loop at each temporal point. A section of this sphere is plotted
in  figure 2C, which shows the absolute value of the determinant $D(t,x,y,z)$ to the
$1/4$ power as a function of $x$ and y, with $z$ fixed to the value that maximizes
the action density and $t$ fixed midway between the two merons. As claimed, this
section is clearly observed to be circular.

We have obtained analogous results for all four configurations we have studied.
The locus of all the points that can be monopoles or vortices is a three-dimensional 
volume as described above, joining the two meron components. The width of this volume 
joining the merons increases with increasing separation between the merons, and the 
maximum width at the midpoint is approximately 4, 4.5, and 6 for $d$ = 10, 12, and 16
respectively. The only effect of a few cooling sweeps applied to these configurations 
is a small change in  the positions and widths of the merons, and the resulting 
monopole and vortex content is the same as described above relative to the new
positions of the merons.

Finally, it is interesting to consider how the monopole and vortex content we
found for the meron pair connects with the two limiting cases discussed before,
an isolated instanton and a single meron. If we think of an instanton as a meron pair
and dissociate it into two separated merons (keeping one of them fixed at the origin),
we see that the  monopole content of the instanton, a point at the maximum, becomes 
a loop of the form shown in figure 2B joining both merons.  As we continue to 
separate both components, this loop becomes larger, and when one of the components 
approaches infinity, this loop joins the meron at the origin with the meron at
infinity, the result we have already discussed for the single meron solution. 
The same argument is valid for the vortex content. As we separate the merons, a 
vortex surface joins both components, and as the separation approaches infinity, 
in the vicinity of one meron the surface locally looks like the planar vortex 
surface of a single meron.

\section{Conclusions}

We have investigated the monopole and vortex content of a meron pair by
calculating the points at which the gauge transformation fixing the gauge to the
Laplacian Center Gauge is ill-defined. Threefold degeneracy of the lowest
eigenvalues of the Laplacian  allows the choice of different
pairs of vectors to define the Laplacian Center Gauge, giving rise to
different pictures for the monopole and vortex content of the configuration.
The determinant of these three eigenfunctions at each lattice point defines the
locus of all the points that can be monopoles or vortices. This locus is a
three-dimensional volume joining both merons, and at each time plane between
the merons, the locus is the surface of a sphere with its center on the line
connecting the merons, and with a diameter given by the width of the curve
shown in figure 2B. One particular choice of degenerate eigenvectors has a
monopole line joining both merons in the $xt$ plane for the first vector
and in the $yt$ plane for the second vector.  The corresponding vortex surface
looks like an ellipsoid of revolution around the $t$ axis in the
$x,y,t$ coordinates. Many other choices of two combinations of the three
degenerate eigenvectors are possible, but all monopole lines and vortex
surfaces must lie in the volume where the determinant vanishes. In
particular, this implies that at the position of the meron, the vortex must
always be in a purely spatial plane and can never be in a space-time plane.

We have also investigated the vortex content of the meron pair by
calculating Wilson loops in all Cartesian planes containing the merons. This
calculation showed that as in the case of an isolated meron, each meron in a 
pair behaves like a source or sink for flux in  non-trivial elements of the 
center of the group for all six planes defined by the Cartesian axes and thus 
corresponds to the intersection of orthogonal pairs of vortices. Thus the Wilson 
loops imply that in addition to a vortex at the position of a meron in a purely 
spatial plane, there must also be a second vortex in a space-time plane.

Although these two complementary investigations have provided interesting 
insight into the vortex structure of a meron pair, it is clear
from comparing the results that Laplacian Center Gauge  fixing is not a
sufficiently powerful tool to reveal the full structure of intersecting
vortices. Whereas Wilson loops clearly imply the intersection of both spatial and
space-time vortices at the merons, Laplacian Center Gauge fixing only finds vortex
surfaces joining the two merons that intersect the merons in spatial planes.
We note that the high symmetry of the background field produces a
highly atypical situation including, for example, the intersection of monopole
loops and a high degeneracy of equivalent solutions. It is possible that the
introduction of a small perturbation would not only remove the intersections
and degeneracy, but also produce a more generic situation of intersecting
vortices.  If this is not the case, more powerful techniques will be required 
to fully analyze the vortex structure.

Finally, looking at the combination of the  results we obtain for the meron
pair from Wilson loops and Laplacian Center Gauge fixing, it is reasonable to
conclude that in a pair as well as in isolation, a meron is a localized source of
monopole trajectories and a localized object with topological charge $\frac{1}{2}$
carrying center flux in six orthogonal space-space and space-time planes.

\end{document}